\documentclass[9pt]{article}
\usepackage{latexsym}
\usepackage{graphicx}
\usepackage{amsfonts,amssymb}

\begin{document}
\title{\textbf{Upper Bound of Relative Error of Random Ball Coverage for Calculating Fractal Network Dimension}}
\author{Yanqing Hu, Zengru Di\footnote{Author for correspondence: zdi@bnu.edu.cn}\\
\\\emph{ Department of Systems Science, School of Management,}
\\\emph{Center for Complexity Research, }\\ \emph{Beijing Normal University, Beijing 100875, P.R.China}}
\maketitle

\begin{abstract}
Least box number coverage problem for calculating dimension of
fractal networks is a NP-hard problem. Meanwhile, the time
complexity of random ball coverage for calculating dimension is very
low. In this paper we strictly present the upper bound of relative
error for random ball coverage algorithm. We also propose
twice-random ball coverage algorithm for calculating network
dimension. For many real-world fractal networks, when the network
diameter is sufficient large, the relative error upper bound of this
method will tend to $0$. In this point of view, given a proper
acceptable error range, the dimension calculation is not a NP-hard
problem, but P problem instead.
\end{abstract}



\section{Introduction}
The structure of complex systems across various disciplines can be
abstracted and conceptualized as complex networks of nodes and links
to which many quantitative methods can be applied so as to extract
any characteristics embedded in the system\cite{Albert reviwe,Newman
reviwe,new reviwe}. There exist many types of networks and
characterizing their topology is very important for a wide range of
static and dynamic properties. Recently, C. Song \textit{et
al.}\cite{Song fractal 1,Song fractal 2} applied a least box number
covering algorithm to demonstrate the existence of self-similarity
in many real networks. They also studied and compared several
possible least box number covering algorithms, by applying them to a
number of model and real-world networks\cite{Song fractal
algorithm}. They found that the least box number covering
optimization is equivalent to the well-known vertex coloring
algorithm which is a NP-hard problem. It implied that any least box
number covering algorithms are heuristic algorithms. Can we avoid
the NP-hard problem and simplify the network fractal dimension
calculation? In this paper we modify the random sequential
box-covering algorithm\cite{Ball coverage} and presented a random
ball coverage algorithm. The relative error smallest upper bound
(Supremum) is given theoretically. The simulation experiments shows
that no matter how large the network diameter is, this upper bound
tends to $0.36$. So we develop another algorithm for calculating
dimension which employ only two random ball coverages. We also yield
the smallest relative error upper bound of this algorithm, which
tend to $0$ when the network diameter is large enough. In this point
of view, for a proper acceptable error range, the random ball
covering algorithm is equivalent to the least box number covering
algorithm in statistic sense when a network is large enough. There
is no need to focus on the least box number covering optimization
problem when we want to calculate dimension of a large diameter
network.

\section{Strict upper bounds of random ball coverage}
The least box number coverage\cite{Song fractal algorithm} and
random ball coverage were defined as following. For a given network
$G$, a box with diameter $r$ is a set of nodes where all distances
$d_{ij}$ between any two nodes $i$ and $j$ in the box are smaller
than $r$. The least box number coverage is the box coverage with the
minimum number of boxes required to cover the entire network $G$. In
order to correspond to the fractal network definition\cite{Song
fractal algorithm} we use 'open' ball to cover the network. So our
random ball coverage is little difference with the random sequential
box-covering algorithm\cite{Ball coverage}. A ball with radius $r$
and center node $c$ is the set of nodes which satisfy the shortest
path length from the center $c$ to each of them is smaller than $r$.
The random ball coverage with radius $r$ as: at each step, we
randomly choose a node which has not been covered as a center, and
cover all the nodes within the distance $r$ to the center. The
process is repeated until all the nodes in the network were covered.

Theorem $1$: $L(2r)\leq$ $B(r)\leq L(r)$, where $L(r)$ is the number
of boxes in a lest box number coverage with diameter $r$ and $B(r)$
denotes the number of balls in a random ball coverage with radius
$r$.

Proof:

$\because\ \ B(\frac{r}{2})$ can be regarded as the number of boxes
in a random box number coverage with diameter $r-1$.

$\therefore\ L(r)\leq L(r-1)\leq B(\frac{r}{2})$

Suppose $L_{1},L_{2},\cdots,L_{m}$ denote all the boxes in a lest
box number coverage, where $m=L(r)$, and
$L_{1}=\{n_{11},n_{12},\cdots,n_{1k_{1}}\},
L_{2}=\{n_{21},n_{22},\cdots,n_{2k_{2}}\},\cdots,\\
L_{m}=\{n_{m1},n_{m2},\cdots,n_{mk_{m}}\}$.
Then we have $L_{i}\bigcup L_{j}=\Phi$ for all $i\neq j, i\leq m,
j\leq m$, where $\Phi$ denotes empty set.

According to the above definition of random ball coverage with
radius $r$, without loosing any generality suppose the center of the
first ball in a random ball coverage process is $n_{11}$, then
$L_{1}$ is covered by the first ball and the second random ball's
center must lies out of $L_{1}$. Without loosing any generality we
also can assume  $n_{21}$ is the center of the second ball, then
$L_{2}$ is covered by the second ball. In this way, we can get that:
$B(r)\leq L(r)$.

$\therefore\ B(r)\leq L(r)\leq B(\frac{r}{2})$

$\therefore\ L(2r)\leq$ $B(r)\leq L(r)$

Theorem $2$: Suppose $G$ is a fractal network, the available box
diameter range is $\{m,m+1,\ldots,R\}$ and we employ linear lest
squares regression to get the dimension. Then the smallest upper
bound of the relative error of dimension calculated by random ball
coverage is

$e(R,m)=\frac{\log2[k\log\frac{R!}{m!}-(R-m+1)\log\frac{(m+k-1)!}{m!}]}{(R-m+1)\sum_{i=m}^{R}{(\log{i})^{2}}-(\log\frac{R!}{m!})^2}$

 where, $k\in\{1,2,\cdots,R-m+1\}$ and satisfy:
$\log(m+k-1)\leq\frac{\log\frac{R!}{m!}}{R-m+1}\leq\,\log(m+k)$
Proof:

$\because$ network $G$ is fractal in range $\{m,m+1,\ldots,R\}$.

$\therefore$ there must exist a proper $b$ for any
$r\in\{m,m+1,\ldots,R\}$ such that
\begin{equation}
\log_{a}L(r)=-s\log_{a}r+b \label{line1}
\end{equation}
\begin{equation}
\log_{a}L(2r)=-s\log_{a}r+b-s\log_{a}2 \label{line2}
\end{equation}

Then
\begin{equation}
\log_{a}B(r)=-s\log_{a}r+b-\theta_{r}s\log_{a}2, \theta_{r}\in[0,1]
\end{equation}
where, $s$ is the dimension of the network.

Suppose
\begin{equation}
Log_{a}B(r)=-\hat{s}Log_{a}r+\hat{b}+\delta_{r}\label{line3},
\end{equation}
$A=\left(
     \begin{array}{cccc}
       \log_{a}m & \log_{a}(m+1) & \cdots & \log_{a}R \\
       1 & 1 & \cdots & 1 \\
     \end{array}
   \right)
$

$b=\left(
     \begin{array}{cccc}
       \log_{a}B(m) & \log_{a}B(m+1) & \cdots & \log_{a}B(R) \\
     \end{array}
   \right)
$

Then $(\hat{s},\hat{b})'=(AA')^{-1}Ab'$

and
$\hat{s}=-s-s\frac{\log2[(R-m+1)\sum_{i=m}^{R}\theta_{i}\log{i}-\sum_{i=m}^{R}\theta_{i}\cdot\log\frac{R!}{m!}]}{(R-m+1)\sum_{i=m}^{R}{(\log{i})^{2}}-(\log\frac{R!}{m!})^2}$

$\therefore$ the relative error is
$\epsilon=\frac{\log2[(R-m+1)\sum_{i=m}^{R}\theta_{i}\log{i}-\sum_{i=m}^{R}\theta_{i}\cdot\log\frac{R!}{m!}]}{(R-m+1)\sum_{i=m}^{R}{(\log{i})^{2}}-(\log\frac{R!}{m!})^2}$

\begin{figure}
\center
\includegraphics[width=8cm]{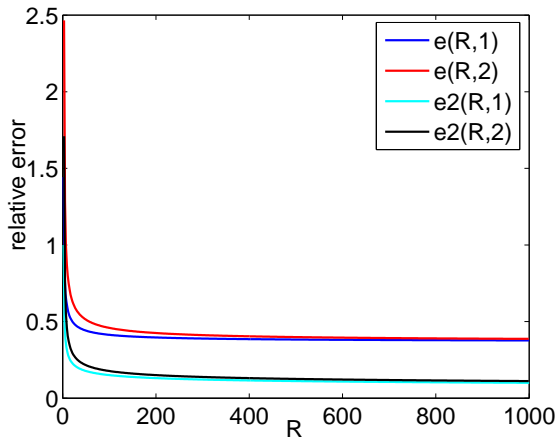}
\caption{Relative error plot. $R$ denotes the diameter of a network,
$e,e2$ denote the relative errors of random ball coverage algorithm
and twice-random ball coverage
algorithm.}\label{Relative_error_plot}
\end{figure}

Employing linear programming, the upper bound of the relative error

is:\,\,\,$e(R,m)=\frac{\log2[k\log\frac{R!}{m!}-(R-m+1)\log\frac{(m+k-1)!}{m!}]}{(R-m+1)\sum_{i=m}^{R}{(\log{i})^{2}}-(\log\frac{R!}{m!})^2}$

where, $k\in\{1,2,\cdots,R-m+1\}$ and satisfy:
$\log(m+k-1)\leq\frac{\log\frac{R!}{m!}}{R-m+1}\leq\,\log(m+k)$
Fig.\ref{Relative_error_plot} show the relationship among $R, m,e$.

It is not easy to get any conclusion directly about the upper bound
from the above expression. So we have done some numerical
calculations. It seems that for a given $m$ when $R$ becomes large,
it has a limit. For instance, when $m$=1 or 2, its limit is about
$0.36$. It implies that no matter how large the network diameter is,
the relative error never be lower than $0.36$. In fact, we could
only use two points $B(m)$ and $B(R)$ to estimate dimension $s$
which is named twice-random ball coverage algorithm. The
corresponding upper bound of relative error will be tend to $0$ when
the fractal network diameter tend to infinite.

Theorem $3$: Suppose $G$ is a fractal network, the available box
diameter range is $\{m,m+1,\ldots,R\}$. We just use $B(m)$ and
$B(R)$ to estimate network dimension. Then the estimated dimension
value is
$\hat{s}=\frac{\log_{a}B(m)-\log_{a}B(R)}{\log_{a}R-\log_{a}m}$ and
the smallest relative error upper bound is $\log_{\frac{R}{m}}2$.

Proof:

Obviously,
$\max\hat{s}=s_{1}=\frac{\log_{a}L(m)-\log_{a}L(2R)}{\log_{a}R-\log_{a}m}=s\log_{\frac{R}{m}}\frac{R}{2m}$
and

$\min\hat{s}=s_{2}=\frac{\log_{a}L(2m)-\log_{a}L(R)}{\log_{a}R-\log_{a}m}=s\log_{\frac{R}{m}}\frac{2R}{m}$
, then
\begin{equation}
e2(R,m)=\mid\frac{\hat{s}-s}{s}\mid\leq \max\{\mid\frac{
s_{1}-s}{s}\mid,\mid\frac{
s_{2}-s}{s}\mid\}=\log_{\frac{R}{m}}2\label{upper bound}
\end{equation}

More details are shown in Fig.\ref{Relative_error_plot}.

Employing random ball coverage algorithm (twice-random ball coverage
algorithm), we get the fractal dimension of the world-wide web is
$s=4.16(3.61)$, $(R=$diameter, $m=1)$ ,which is corresponding to the
dimension obtained by C. Song \textit{et al.}(dimension is
$s=4.1$)\cite{Song fractal 1}. From our empirical results, we find
for many networks, $R=$diameter and $m=2$ is more reasonable. When
$m=2$ we get WWW network dimension is $4.48(4.51)$. Sometimes, the
available box diameter $R$ is not sufficient enough, we can
calculate $B(r)$ many time and get the dimension. We also test this
method in the $43$ cellular networks\cite{metabolic}. For each
network $(R$=diameter, $m=2)$ and each $B(r)$, we perform random
coverage $100$ times. Then we get the average dimension of the whole
cellular networks is $s=3.54(3.58)$ which is perfect corresponding
to the dimension obtained by C. Song \textit{et al.}(dimension is
$s=3.5$)\cite{Song fractal 1} and W. Zhou \textit{et al.} (dimension
is $s=3.54\pm0.27$)\cite{Edage coverage}. Because we calculate each
$B(r)\ 100$ times, for any one of the $43$ cellular networks, we can
get $100$ different dimensions. For each cellular network we can get
an average variance. The maximum average variance of $43$ cellular
network dimensions is $0.042(0.062)$, the average variance
$0.018(0.023)$. If we use the network dimension which is obtained by
$100$ times calculation to substitute its real fractal dimension $s$
in our above discussion, we get the maximum relative error of $100$
time calculations of the $43$ cellular networks is $0.15(0.20)$. The
average maximum relative error is $0.045(0.11)$ and the average
relative error is $0.030(0.034)$. The relative errors of empirical
results are far less than the theoretical upper bounds respectively.
The interesting thing is that, in our theoretical discussion, the
upper bound of twice-random ball coverage is less than the upper
bound of random ball coverage algorithm. But the empirical results
always show the random ball coverage algorithm is better than
twice-random ball coverage algorithm. So, we think the random ball
coverage algorithm is better than twice-random ball coverage
algorithm in practice. Moreover, our theorems also can be used to
estimate a network's diameter.

\section{Conclusion and discussion}
In this paper, we strictly present the upper bound of the relative
error of random ball coverage method in fractal network dimension
calculation. And we also yield a simple relative error upper bound
$\log_{\frac{R}{m}}2$ of twice-random ball coverage method. For many
real-world networks, when the network diameter is sufficient enough
this kind of relative error upper bound will tend to $0$. Therefore,
if the network is sufficient enough, twice-random ball coverage is
equivalent to the leat box number coverage in fractal dimension
calculation and calculating fractal network dimension is not a
NP-hard problem. For the networks which is not sufficient enough, we
can calculate random ball number many times and get the dimension,
which is also very effective and accuracy.

The above discussions can lead another problem naturally. We also
can define random full box coverage. A full box with diameter $r$ is
a set of nodes, such that any other nodes out of the box is added to
the box will make the box diameter larger or equal to $r$. The
random full box coverage algorithm with diameter $r$ can be defined
as\cite{Song fractal algorithm}: at each step we randomly choose a
uncovered node $p$ as the first node of the box, and select the
uncovered nodes to the box until the box become full. We guess the
random full box coverage algorithm is equivalent to the least box
number covering algorithm in statistic sense. In the future we will
do some deep researches about this problem.

\section*{Acknowledgement}
The authors want to thank Chaoming Song, Qiang Yuan for provide some
useful information. This work is partially supported by 985 Projet
and NSFC under the grant No.$70431002$, No.$70771011$ and
No.$70471080$.

\end{document}